\documentclass[twocolumn,showpacs,preprintnumbers,amsmath,amssymb]{revtex4}

\usepackage{graphicx}
\usepackage{amssymb}
\usepackage{dcolumn}

\begin{document}
\title{Condensation phase transitions of symmetric conserved-mass
aggregation model on complex networks}

\author{Sungchul Kwon}
\author{Sungmin Lee}
\author{Yup Kim}
\email{ykim@khu.ac.kr} \affiliation{Department of Physics and
Research Institute for Basic Sciences, Kyung Hee University, Seoul
130-701, Korea}
\date{\today}

\begin{abstract}
We investigate condensation phase transitions of symmetric
conserved-mass aggregation (SCA) model on random networks (RNs)
and scale-free networks (SFNs) with degree distribution $P(k) \sim
k^{-\gamma}$. In SCA model, masses diffuse with unite rate, and
unit mass chips off from mass with rate $\omega$. The dynamics
conserves total mass density $\rho$. In the steady state, on RNs
and SFNs with $\gamma>3$ for $\omega \neq \infty$, we numerically
show that SCA model undergoes the same type condensation
transitions as those on regular lattices. However the critical
line $\rho_c (\omega)$ depends on network structures. On SFNs with
$\gamma \leq 3$, the fluid phase of exponential mass distribution
completely disappears and no phase transitions occurs. Instead,
the condensation with exponentially decaying background mass
distribution always takes place for any non-zero density. For the
existence of the condensed phase for $\gamma \leq 3$ at the zero
density limit, we investigate one lamb-lion problem on RNs and
SFNs. We numerically show that a lamb survives indefinitely with
finite survival probability on RNs and SFNs with $\gamma >3$, and
dies out exponentially on SFNs with $\gamma \leq 3$. The finite
life time of a lamb on SFNs with $\gamma \leq 3$ ensures the
existence of the condensation at the zero density limit on SFNs
with $\gamma \leq 3$ at which direct numerical simulations are
practically impossible. At $\omega = \infty$, we numerically
confirm that complete condensation takes place for any $\rho > 0$
on RNs. Together with the recent study on SFNs, the complete
condensation always occurs on both RNs and SFNs in zero range
process with constant hopping rate.

\end{abstract}
\maketitle

\section{Introduction}
Nonequilibrium condensation phase transitions from fluid phase
into condensed phase have been observed in a variety of phenomena
ranging from traffic flow to polymer gels
\cite{zrp,E2,LEC,M,DM,NSL,zrpdirectnet,Z,W,S,MRRGI,F,MKB}. In the
steady state, a finite fraction of total particles condenses on a
single site in condensed phase when the total particle density
$\rho$ is increased beyond a certain critical value $\rho_c$. In
fluid phase below $\rho_c$, the particle number of each site
fluctuates around $\rho$ without the condensation.

Various kinds of nonequilibrium mass transport models exhibit the
condensation transitions or only condensation. The simplest and
well-known model is zero-range process(ZRP) in one-dimension
\cite{zrp}. In ZRP, many identical particles occupy sites on a
lattice. Each site may contain an integer number of particles and
one of these particles can hop to one of the nearest neighboring
sites with a rate that depends on the number of particles at the
site of departure. The chipping (single-particle dissociation) and
aggregation processes of ZRP describe various condensations such
as jamming of traffic \cite{E2}, bunching of buses \cite{LEC},
coalescence of shaken steel balls \cite{M} and condensation of
edges in networks \cite{DM}. Recent studies of ZRP on scale-free
\cite{NSL} and directed networks \cite{zrpdirectnet} reveal the
conditions under which condensation takes place.

Another important class of condensation transitions emerges when
the diffusion of the whole particles of a single site is involved
in addition to the chipping and aggregation. These processes arise
in a variety of phenomena such as polymer gels \cite{Z}, the
formation of colloidal suspensions \cite{W}, river networks
\cite{S,MRRGI} and clouds \cite{F}. Recently studied
conserved-mass aggregation(CA) model is the simplest one
incorporating diffusion, chipping and aggregation upon contact
\cite{MKB,exactsca}. In one dimensional CA model, the mass $m_i$
of site $i$ moves either to site $i-1$ or to site $i+1$ with unit
rate, and then $m_i \rightarrow 0$ and $m_{i\pm1} \rightarrow
m_{i\pm1}+m_i$. With rate $\omega$, unit mass chips off from site
$i$ and moves to one of the nearest neighboring sites; $m_i
\rightarrow m_i -1$ and $m_{i\pm1}\rightarrow m_{i\pm1}+1$. As
total masses are conserved, the conserved density $\rho$ and
$\omega$ determine the phase of CA model. The condensation
transition arises via the competition between diffusion and
chipping processes. The diffusion of masses tends to produce
massive aggregates and consequently creates more vacant sites. The
chipping of unit mass tends to prevent the formation of
aggregates, so that it leads to a replenishment of the lower end
of mass distribution.

The single site mass distribution $P(m)$, i.e., the probability
that a site has mass $m$ in the steady state, was shown to undergo
phase transitions on regular lattices \cite{MKB}. For a fixed
$\omega$, as $\rho$ is varied across the critical value $\rho_c
(\omega)$, the behavior of $P(m)$ for large $m$ was found to be
\cite{MKB}
\begin{equation}
 P(m)\sim \begin{cases}
          e^{-m/m^{*}} &\rho<\rho_c (\omega), \\
          m^{-\tau} & \rho=\rho_c (\omega), \\
          m^{-\tau}+ \text{\rm infinite\,\,\, aggregate} &\rho>\rho_c
          (\omega),
          \end{cases}
\end{equation}
where $\rho_c$ is given as $\rho_c (\omega) = \sqrt{\omega+1}-1$.
$\rho_c$ and $\tau$ are shown to be independent of spatial
dimension $d$ \cite{exactsca}. The tail of the mass distribution
changes from exponential to an algebraic decay as $\rho$
approaches $\rho_c$ from below. As one further increases $\rho$
beyond $\rho_c$, this asymptotic algebraic part of the critical
distribution remains unchanged but in addition an infinite
aggregate forms. This means that all the additional mass
$(\rho-\rho_c)L^d $ condenses onto a single site and does not
disturb the background critical distribution. The $\omega =
\infty$ case corresponds to ZRP with constant chipping rate, and
then there is no condensation transitions on regular lattices. The
critical exponent $\tau$ is same everywhere on the critical line
$\rho_c (\omega)$. Recent studies showed that $\tau$ depends on
the symmetry of movement and constraints of diffusion rate
\cite{RK,ca-mass}. In what following, we only consider the
symmetric CA (SCA) model where diffusion and chipping directions
are unbiased \cite{MKB}.

As mean field theory correctly predicts the phase diagram and
$\tau$ of SCA model in one dimension \cite{MKB}, the nature of the
condensation transitions should remain unchanged in any higher
dimensions. However recent studies of various dynamics on network
structures showed that the structure of networks leads to more
rich and intriguing behavior different from that predicted by the
standard mean-field theory on regular lattices
\cite{DM,NSL,mendes03}. For example, ZRP on SFNs was shown to
exhibit condensation even at constant hopping rate which is the
case of an infinite chipping rate of SCA model \cite{NSL}.

In this paper, we investigate the effect of network structures on
the condensation transitions of SCA model using random networks
(RNs) and scale-free networks (SFNs). As we shall see, on RNs and
SFNs with the degree exponent $\gamma >3$, SCA model undergoes the
same type of condensation transitions as those in regular lattice
across a critical line $\rho_c (\omega)$ in $\rho$-$\omega$ plane
with the exponent $\tau = 5/2$. However on SFNs with $\gamma \leq
3$ where one or several nodes, so-called hub nodes, have a finite
fraction of links, the fluid phase completely disappears and the
condensation with exponentially decaying background mass
distribution takes place for any non-zero density. The out-line of
this paper is as follows. In Sec. II, we introduce SCA model on
complex networks. The condensation transitions on RNs and SFNs are
discussed in Sec. III and IV. To understand the condensation on
SFNs with $\gamma \leq 3$, we discuss lamb-lion problems on SFNs
in Sec. V. In Sec. VI and VII, we discuss SCA model at
$\omega=\infty$ and the effect of diffusion of masses on average
mass distribution on degrees respectively. Finally we summarize
our results in Sec. VIII.

\section{SCA Model on networks}

We consider a network with $N$ nodes and $K$ links. The degree
$k_i$ of a node $i$ is defined as the number of its links
connected to other nodes. The average degree of a node $<k>$ is
given as $<k>=2K/N$. The degree distribution $P(k)$ is a Poisson
distribution for RNs and a power-law distribution of $P(k)\sim
k^{-\gamma}$ for SFNs. Each node may have an integer number of
particles, and the mass of a node is defined as the number of
particles at the node. Initially $M$ particles randomly distribute
on $N$ nodes with given conserved density $\rho = M/N$. Next a
node $i$ is chosen at random and one of the following events is
occurred:

(i) Diffusion: With unit rate, the mass $m_i$ moves to the
randomly selected nearest neighboring node $j$. If the node $j$
has already mass $m_j$, then the aggregation takes place; $m_i
\rightarrow 0$ and $m_j \rightarrow m_j + m_i$.

(ii) Chipping: With rate $\omega$, unit mass (a single particle)
at node $i$ chips off and moves to the randomly selected nearest
neighboring node $j$; $m_i \rightarrow m_i -1$ and $m_j
\rightarrow m_j + 1$.

As mentioned in Sec. I, diffusion and chipping processes compete
each other so a condensation transition physically depends on two
external parameters $\rho$ and $\omega$. At $\omega =0$, however,
only diffusion and aggregation take place so masses always
condense onto a single site (complete condensation). On the other
hand, the $\omega=\infty$ case corresponds to ZRP with constant
chipping rate. It was shown that ZRP on scale-free networks
exhibits complete condensation when mass-dependent chipping rate
$u(m)$ is given as $u(m)\sim m^\delta$, but $\delta < \delta_c$,
where $\delta_c$ is some threshold value \cite{NSL}. Hence the SCA
model of $\omega = \infty$ is just the $\delta =0$ case of ZRP so
the complete condensation always aries at $\omega=\infty$. As we
shall see in the following section, the complete condensation also
takes place on random networks at $\rho
>0$.

For the construction of SFN, we use a static model \cite{SNUGroup}
instead of preferential attachment algorithm \cite{DM}. In static
model, it is desired to use large $<k>$ to construct fully
connected networks. In simulations, we use $<k>=4$.

\section{SCA model on RNs and SFNs with $\gamma >3$}

We consider SCA model on RNs and SFNs with $\gamma>3$ having nodes
$N=10^4$ and links $K=2\times 10^4$ for $\omega \neq \infty$. We
define the single node mass distribution $P(m,t)$ at time $t$,
i.e., the probability that a node has mass $m$ at $t$ as
\begin{equation}
P(m,t)= \frac{1}{N} n(m,t)=\frac{1}{N} \sum_{i=1}^N \delta_{m_i
(t),m} \;\; ,
\end{equation}
where $n(m,t)$ is the number of nodes with mass $m$ at time $t$
and $m_i (t)$ is the mass of node $i$ at $t$.
From the normalization condition of $P(m,t)$, i.e., $\sum_m P(m,t)
= 1$, all $P(m,t)$ should reach the stationary state at the same
time. For instance, $P(m=0,t)$ is given as $P(0,t)=1-\sum_{m \neq
0} P(m,t)$. Hence, if $P(0.t)$ saturates to its steady state value
$P(0)$ after some time $\tau_0$, then $P(m,t)$ for any mass $m$
also should reach its steady state value $P(m)$ after the time
$\tau_0$. So the characteristic time $\tau_m$ at which $P(m,t)$
saturates is expected to be independent of mass $m$, and
\begin{eqnarray}\label{eq:tau}
\tau_0 = \tau_1 = \cdots =\tau_{m^*} \;\;,
\end{eqnarray}
where $m^*$ is the maximal mass. In Fig. 1, we plot $P(m,t)$ of
$m=1$ and $10$ for $\omega=1.0$ and $\rho=0.1$ on various
networks. The vertical dotted line denotes the characteristic time
$\tau_m$, which supports Eq. (\ref{eq:tau}).

We measure the steady state distribution $P(m)$ by averaging
$P(m,t)$ after $\tau_m$. Fig. 2 shows the plots of $P(m)$ versus
$m$ for RN and SFN with $\gamma = 4.3$ with $\omega=1.0$. The
critical density $\rho_c$ is $\rho_c (\omega=1) = 0.31(3)$ for RN
and $0.24(3)$ for SFN with $\gamma=4.3$ respectively. In fluid
phase ($\rho <\rho_c$), $P(m)$ exponentially decays for large mass
in both RN and SFN with $\gamma = 4.3$. At $\rho=\rho_c$, $P(m)$
algebraically decays on both RN and SFN with $\gamma=4.3$ as $P(m)
\sim m^{-\tau}$ with $\tau = 2.38(3)$ for RN and $\tau = 2.33(2)$
for SFN with $\gamma=4.3$ without condensations. The values of
$\tau$ of both networks also agree well with $\tau=5/2$ of regular
lattices \cite{MKB,exactsca}. In condensed phase ($\rho=3.0 >
\rho_c)$, the access mass of $(\rho-\rho_c)N \approx 2.7 \times
10^4$ condenses on a single node in the both networks without
changing the background critical distribution. To map out phase
diagrams for RN and SFN with $\gamma=4.3$, we measure $\rho_c$ and
$\tau$ for $\omega= 0.1, 1.0$ and $10$, and find that $\tau$ is
the same within error on the critical line. Fig. 3 shows the phase
diagrams of RN and SFN with $\gamma=4.3$. The solid line denotes
the critical line, $\rho_c = \sqrt{\omega+1}-1$ of regular
lattices \cite{MKB}. On regular lattices, the $\rho_c (\omega)$ is
independent of dimensionality. However on networks, $\rho_c
(\omega)$ depends on the underlying network structures. We also
confirm that SFN with $\gamma=3.5$ undergoes the same type of
condensation transitions. Therefore we conclude that SCA model on
RNs and SFNs with $\gamma>3$ undergoes the same type of
condensation transitions from fluid phase into condensed phase for
$0<\omega<\infty$ as those on one-dimensional regular lattice.
However the critical line depends on the underlying network
structures.

\section{SCA model on SFN with $2 \leq \gamma \leq 3$}

For $\omega \neq \infty$, we perform simulations on SFNs with
$\gamma = 2.4$ and $3$ with $N=10^4$ nodes and $K=2\times 10^4$
links. We plot $P(m,t)$ of $\gamma=2.4$ in Fig. 1 (c) for example,
and average $P(m,t)$ after $\tau_m$ for the steady state $P(m)$.
As mentioned in the previous section, $\tau_m$ is same for all
masses.

The mass distribution $P(m)$ for $\gamma \leq 3$ show quite
different behavior from that for $\gamma > 3$. Fig. 4 shows $P(m)$
on SFNs with $\gamma= 2.4$ and $3$ for $\rho=0.2$ and $3.0$ with
$\omega=1$. As in the condensed phase of RNs and SFNs with
$\gamma>3$, there is a condensation of mass $m^* \approx \rho N$.
However $P(m)$ exponentially decays for large mass rather than
power-law. We also measure $P(m)$ for various $\omega$ and $\rho$,
and confirm the same behavior of $P(m)$. Therefore we conclude
that for $\omega \neq \infty$, SCA on SFNs with $\gamma \leq 3$
exhibits an infinite aggregation with an exponential background
mass distribution. We call this phase incomplete condensed phase
to reflect exponential mass distributions (Fig. 4 (c)).
Intriguingly the two features of the fluid and the condensed phase
on regular lattice coexist on SFNs with $\gamma \leq 3$. Such a
behavior was found in the lattice gas model proposed to describe
the distribution of droplets in the fragmentation process
following a nuclear collision \cite{campi}.

 The nonexistence of the fluid phase for
$\gamma \leq 3$ mainly comes from the hub structure of SFNs with
$\gamma \leq 3$ where one or several nodes have a finite fraction
of total links \cite{DM}. Due to the hub structure, all masses
move to a hub node by both diffusion and chipping processes unlike
SFNs with $\gamma >3$ where the chipping tends to split masses and
prevents aggregations. Hence there are no processes to prevent the
formation of infinite aggregation for $\gamma \leq 3$. The
difference from ZRP is that the diffusion process moves an
infinite aggregation of mass $m_{hub}$ at a hub node to the
others. Then the chipping distributes small masses of $m_{hub}$
onto neighboring nodes. In this way, nodes with small degree can
have mass so the resultant $P(m)$ is exponential.

It is hard to show the existence of infinite aggregation for the
limit of $\rho \rightarrow 0$ numerically with finite size
networks. Instead we investigate annihilating random walks of two
particles on SFNs (lamb-lion problem) for the condition of the
existence of the fluid phase on SFNs. For $\rho \rightarrow 0$
limit, we assume an infinite aggregation and neglect the presence
of other masses. Then unit mass chips off from the aggregation,
and diffuses around on the networks. For the existence of an
infinite condensate in the steady state, the two masses should
aggregate again in finite time interval. If not, unit mass
continuously chips off from the infinite aggregation, which will
finally disappear. Hence if the unit mass (lamb) survives
indefinitely without meeting the infinite aggregation (lion), then
the fluid phase exists at zero density limit. Otherwise, the
condensed phase exists for any $\rho>0$. In next section, we
investigate a lamb-lion problem on RNs and SFNs.

\section{Lamb-lion problem on networks}
We consider one lamb and one lion problem on networks. If a lamb
meets a lion on the same node, then it dies. Hence the interesting
quantity is the survival probability $S(t)$ of a lamb at time $t$.
On $d$ dimensional regular lattice, $S(t)$ depends on $d$ as
follows \cite{BG,Feller,Weiss,KR}.
\begin{equation}
S(t) \sim \begin{cases}
           t^{d/2-1} \;\;\;\;\; \text{for $d<2$} \;\;, \\
           1/\ln{t} \;\; \;\;\;\text{for $d=2$} \;\;, \\
           \text{constant for $d>2$} \;\;.
           \end{cases}
\end{equation}
On networks, the steady state distribution of a random walker is
proportional to the degree $k_i$ of node $i$ \cite{NR} as
\begin{equation}
\label{pi}
 P^{\infty}_i = k_i /\sum_{j=1}^{N} k_j .
\end{equation}
Hence the probability of finding two walkers at the same node
should depend on the second moment $<k^2>$. While $<k^2>$ is
finite for RNs and SFNs with $\gamma >3$, it diverges for SFNs
with $\gamma \leq 3$. So RNs and SFNs with $\gamma >3$ correspond
to infinite dimensional homogeneous regular lattice in macroscopic
scale, while SFNs with $\gamma \leq 3$ are generically
inhomogeneous. Therefore $S(t)$ is expected to be constant in the
thermodynamic limit on RNs and SFNs with $\gamma >3$. It means the
existence of the fluid phase for the limit of $\rho=0$ for any
$\omega \neq \infty$ as shown in Fig. 3. On the other hand, for
the incomplete condensed phase of $\gamma \leq 3$, $S(t)$ should
decay to zero for $\gamma \leq 3$.

We measure $S(t)$ of a lamb on RNs and SFNs with various $\gamma$
for several values of $N$ from $10^3$ to $10^6$ and up to $10^6$
time steps. Initially a lamb and a lion are placed on randomly
selected two nodes. In Fig. 5, we plot $S(t)$ for $\gamma =4.3$
(a) and $2.4$ (b) for various system size $N$. Interestingly
$S(t)$ decays exponentially in both SFNs for any $N$. For example,
$S(t)$ of $\gamma=4.3$ and $N=10^6$ exponentially decays with very
large characteristic time $\tau$ order of $10^6$ over the whole
time interval. We also confirm the same exponential decay of
$S(t)$ on RNs for any $N$. Unlike on regular lattices where the
exponential decay of $S(t)$ comes from finite size effect of
lattices, $S(t)$ on networks shows such a exponential decay at
very early time. The exponential decay of $S(t)$ mainly comes from
the small world nature of networks \cite{DM}, i.e., the average
path length between a pair of nodes increases logarithmically with
$N$. As $S(t)$ shows the exponential decay for any $\gamma$, we
are interested in the average life time $T$ of a lamb rather than
$S(t)$ itself.

We measure the life time $T$ of a lamb on RNs and SFNs with
several $\gamma$ values for $N$ up to $10^6$ and plot $T$ against
$N$ in Fig. 5 (c) and (d). In Fig. 5 (c), $T$ linearly increases
with $N$ for both RN and SFN with $\gamma=4.3$. Assuming $T \sim
N^{\alpha}$, we estimate $\alpha = 1.0(2)$ for RN and SFN with
$\gamma=4.3$ by measuring slopes. However for $\gamma < 3$ of Fig.
5 (d), $T$ exhibits quit different behavior. $T$ of $\gamma =
2.15$ and $2.4$ tend to saturate to some asymptotic value with
decreasing successive slopes. However the characteristic size $N_s
(\gamma)$ at which $T$ begins to saturate increases as $\gamma$
approaches $\gamma =3$. For example, $T$ of $\gamma=2.75$ seems to
algebraically increase up to $N=10^6$. It means that $N_s $ of
$\gamma=2.75$ is already larger than $10^6$. To see the saturation
at $\gamma=2.75$, big size $N$ much larger than $10^6$ is needed
but it is practically very difficult. It is hard to believe that
$T$ diverges in power-law fashion as $\gamma$ approaches $3$
because $T$ of small $\gamma$ is already begin to saturate at
moderate $N$ order of $10^5 \sim 10^6$. Hence $T$ of $\gamma <3$
is expected to saturate to its asymptotic value even though we
cannot see the saturation via simulations as $\gamma$ approaches
$3$ from below. At $\gamma=3$, $T$ also increases algebraically
with $N$. By measuring successive slopes, we estimate
$\alpha=0.85(1)$ for $\gamma=3$. However, as the exponent $\alpha$
continuously varies as $\gamma$ increases to $3$, the value of
$\alpha$ at $\gamma=3$ may have no special meaning. As shown in
Fig. 4 (a) for $\gamma=3$, the condensation already occurs at
sufficiently low density $\rho=0.2$ even for small size $N=10^4$
so $T$ of $\gamma=3$ is believed to saturate in thermodynamic
limit as for $\gamma=2.75$. Therefore we are convinced that the
life time of a lamb is finite in thermodynamic limit of
$N\rightarrow \infty$ for $\gamma \leq 3$. From the behavior of
the life time $T$, the asymptotic behavior of $S(t)$ is
investigated in the following.

As shown in Fig. 5 (a), $S(t)$ decays exponentially as $S(t) = S_o
e^{-t/\tau}$. From the definition of $T = \int^{\infty}_0 t (-
dS(t)/dt) dt$, we have $T \sim \tau$. Hence $\tau$ diverges
linearly on SFNs with $\gamma >3$ and RNs, while it saturates for
$\gamma \leq 3$. The scaling behavior of $\tau$ means that $S(t)$
on RNs and SFNs with $\gamma > 3$ is nonzero constant in
thermodynamic limit of $N \rightarrow \infty$, and exponentially
decays to zero on SFNs with $\gamma \leq 3$. Hence we have
\begin{equation}
\lim_{N \rightarrow \infty} S(N,t)= \begin{cases}
S_o e^{-t/\tau_{\infty}} & (\gamma \leq 3) \\
S_{\infty} & (\gamma > 3)\;\;,
 \end{cases}
\end{equation}
where $\tau_{\infty}$ is the asymptotic value of $\tau$, and
depends on $\gamma$. As $S(t)$ decays exponentially to zero for
$\gamma \leq 3$, two particles initially located on the same node
should meet or aggregate within finite time interval. Hence the
finite life time on SFNs with $\gamma \leq 3$ supports the
numerically expected condensed phase for any $\rho>0$ on SFNs with
$2 \leq \gamma \leq 3$ as mentioned in the previous section. On
the other hand, the finite $S_\infty$ on RNs and SFNs with $\gamma
>3$ implies the existence of the fluid phase at $\rho \rightarrow
0$ limit, which is also consistent with the phase diagrams of Fig.
3.

\section{SCA model at $\omega = \infty$}

At $\omega = \infty$, SCA model corresponds to ZRP with constant
hopping rate on networks. On SFNs with $\gamma>2$, it was shown
that ZRP undergoes complete condensation for any $\rho >0$
\cite{NSL}. In condensation phenomena of ZRP on SFNs, the average
mass at a node with degree $k$, $m_k$ in the steady state exhibits
a discontinuous jump at a hub node. In ZRP with constant chipping
rate for example, $m_k$ linearly increase until $k < k_{hub}$ and
jumps to the value $m^* \approx \rho N$ at $k=k_{hub}$ \cite{NSL}.
The $k_{hub}$ is the degree of the hub node which has the largest
number of links.

To see whether the condensation occurs on RNs in ZRP, we measure
$m_k$ for $\rho=0.4$ and $3.0$ on RNs with $N=10^4$. As shown in
the bottom inset of Fig. 6 (a), $m_k$ increases linearly in $k$
and finally jumps at $k_{hub}$ as in ZRP on SFNs. For a fixed $N$,
the mass of $k_{hub}$, $m_{hub}$ decreases as $\rho$ goes to zero
as shown in the inset of Fig. 6 (a). So it is hard to observe the
condensation at low density for small $N$. Since the complete
condensation of ZRP on networks mainly comes from the
inhomogeneity of degree distribution rather than the competition
between dynamical processes, ZRP on RNs should exhibit the
complete condensation for any $\rho>0$ as on SFNs. At
$\omega=\infty$, we also measure $m_k$ on SFNs with $\gamma \leq
3$ and confirm the complete condensation as shown in Ref.
\cite{NSL}. Together with the recent study on SFNs \cite{NSL}, the
complete condensation always occurs on both RNs and SFNs in zero
range process with constant hopping rate.

The discontinuity of $m_k$ clearly comes from the nature of
chipping process of unit mass. That is, unit mass hops around
according to the steady state distribution of Eq. (5), and then
there is enough net particle current into the hub nodes to form
infinite aggregation. However when the diffusion of the whole mass
of each node turns on, the aggregation at the hub node diffuses
around all nodes. The diffusion of the condensate completely
change the behavior of $m_k$ in the condensed phase of finite
$\omega$. $m_k$ linearly increase up to $k_{hub}$ without any jump
in the condensed phase of $\omega \neq \infty$. In next section,
we numerically and analytically confirm the linearity of $m_k$ in
$k$.

\section{average mass of a node with degree k}

Another interesting quantity in condensation phenomena on networks
is the average mass at a node with degree $k$, $m_k$ in the steady
state. In ZRP with chipping rate $u(m)\sim m^{\delta}$ on SFNs
\cite{NSL}, the complete condensation takes place for $\delta <
\delta_c = 1/(\gamma -1)$. For $\delta < \delta_c$, $m_k$ linearly
increases with $k$ for $k<< k_c$, and algebraically increases as
$k^{1/\delta}$ for $k \geq k_c$. Especially at $\delta=0$, $m_k$
linearly increases until $k < k_{hub}$ and jumps to the value $m^*
\approx \rho N$. The $k_{hub}$ is the degree of the hub node which
has the largest number of links. The same type of behavior on RN
for $\omega =\infty$ is shown in the inset of Fig. 6 (a).

In SCA model, as all masses can perform random walks according to
the steady state distribution $P^{\infty}_i $ of Eq. (5), one can
expect the jump of $m_k$ at $k_{hub}$ in the condensed phase for
$\omega \neq \infty$ in SCA model as in ZRP. However as shown in
the main plot of Fig. 6 (a), $m_k$ increases linearly in $k$ up to
$k_{hub}$ and there is no jump unlike in ZRP. In Fig. 6 (a), we
plot $m_k$ versus $k$ for $\rho=3.0$ and $\omega=1.0$ which
corresponds to the condensed phase of both RN and SFN with
$\gamma=4.3$. We also measure $m_k$ in the fluid phase ($\rho=0.2,
\omega=1.0$) of the both networks (not shown), and confirm $m_k
\sim k$ without any jumps. For SFN with $\gamma=2.4$, $m_k$ also
shows the linearity in $k$ for $\rho=3.0, \omega=1.0$. Our
simulation results imply that the relation $m_k \sim k$ is valid
for any $\rho >0$ in RNs and SFNs with $\gamma \geq2$ for finite
diffusion rate. The diffusion of masses indeed results in the
linearity of $m_k$ in $k$ for any $\omega \neq \infty$.

In the steady state, as masses can perform random walks with a
finite rate, the mass of a hub node ($m_{hub}$) diffuses to
different nodes with the probability of being at node $i$ given as
Eq. (5) \cite{NR}. The inset of Fig. 6 (b) shows the snapshot of
the mass distribution of nodes with degree $k$ for a single sample
at time $t=4\times 10^5$ on SFN with $\gamma=2.4$ for $\rho=3.0$
and $\omega=1.0$. As shown, there is a peak which may be formed at
$k_{hub}$ as in ZRP. However the peak of mass $m^*$ is not always
located at $k_{hub}$ but diffuse around nodes according to
$P^{\infty}_i$ (Fig. 6 (b)). Hence by taking average, the peak
soaks into the average mass $m_k$ unlike in ZRP where all samples
have the peak at $k_{hub}$. To see this more explicitly, we derive
the relation, $m_k \sim k$, based on the assumption that the
diffusion (the random walks of masses) is the only relevant
physical factor to decide $P(m)$ in the steady state.

First, we consider the average total mass $M_k$ of nodes with
degree $k$ defined as
\begin{eqnarray}\label{eq:bigmk}
M_k = \sum_{m=0}^{\infty} m P_{\infty} (m,k) \;\;,
\end{eqnarray}
where $P_{\infty} (m,k)$ is the probability of finding a random
walker with mass $m$ at degree $k$ in $k$-space in the steady
state. As mass distribution $P(m)$ in the steady state is
independent of $k$, we have $P_{\infty} (m,k) = P(m)
P^{\infty}_k$. $P^{\infty}_k$ is the probability of finding a
random walker at a node with degree $k$ on the network. Then using
$P^{\infty}_i$ of Eq.~(\ref{pi}), we write $P^{\infty}_k$ as
\begin{equation}\label{eq:pkinf}
P^{\infty}_k = \sum_{i=1}^N P^{\infty}_i \delta_{k_i,k} = \frac{k
N P(k)}{\sum_{i=1}^N k_i} \;\;,
\end{equation}
where $P(k)$ is a degree distribution. From Eq. (\ref{eq:bigmk})
and (\ref{eq:pkinf}), we have
\begin{equation}
M_k = \left[\frac{k N P(k)}{\sum_{i=1}^N k_i}\right]
\left[\sum_{m=0}^{\infty} m P(m)\right].
\end{equation}
So $M_k = \bar{m} k P(k)/<k> \sim k P(k)$. Then the average mass
of a node with degree $k$ is given as
\begin{equation}
\label{mk}
 m_k = \frac{M_k}{N P(k)} \sim k/N.
\end{equation}
As expected, $m_k$ scales as $m_k \sim k$ for any degree
distribution $P(k)$. As Eq. (\ref{mk}) is valid for any density
$\rho > 0$, $m_k$ does not undergoes the condensation transitions
unlike $P(m)$ of Eq. (1). Therefore the steady state distribution
of $m_k$ is determined by diffusion of masses rather than chipping
of unit mass unlike in ZRP.

\section{Summary }
We investigate the condensation phase transitions of symmetric
conserved-mass aggregation (SCA) model on networks. In SCA model,
masses diffuse with unite rate, and unit mass chips off from mass
with rate $\omega$. SCA model undergoes condensation phase
transitions via the competition between diffusion and chipping
processes \cite{MKB}.

First we consider the $\omega \neq \infty$ case. On random and
scale-free networks of $\gamma
>3$, SCA model undergoes the same type of condensation transitions
from fluid phase into condensed phase as in one dimensional
lattice of Ref. \cite{MKB}. However unlike on regular lattices,
the critical line on the networks depends on the network
structures. On the other hand, on scale-free networks of $\gamma
\leq 3$ where one or several nodes have a finite fraction of
degrees, an infinite aggregation with exponentially decaying
background mass distribution always takes place for any nonzero
density, so no phase transitions occur for $\omega \neq \infty$.
The condensation and exponential mass distribution of small masses
come from the generic inhomogeneity of network structure of SFNs
with $\gamma \leq 3$. However we are not able to numerically show
the existence of the condensation phase at zero density limit for
$\gamma \leq 3$ due to the small world nature of networks
\cite{DM}. Instead we numerically study the survival probability
of a particle in pair annihilating random walks, so-called one
lamb and one lion problem on networks.

For the formation of an infinite aggregation of masses at zero
density limit, unit mass chipped off from the infinite aggregation
should aggregate again with the aggregation within finite time
interval. We numerically show that in thermodynamic limit, the
survival probability $S(t)$ of a lamb (unit mass) is finite on
random networks and scale-free networks (SFNs) with $\gamma >3$,
but exponentially decays to zero with finite life time on SFNs
with $\gamma \leq 3$. Based on the finite life time of a lamb on
SFNs with $\gamma \leq 3$, we indirectly confirm the incomplete
condensed phase for any $\rho >0$ on SFNs with $\gamma \leq 3$.

At $\omega = \infty$, SCA model corresponds to zero-range process
(ZRP) with constant chipping rate \cite{zrp}. ZRP with constant
chipping rate on SFNs was shown to exhibit complete condensation
\cite{NSL}. We also numerically show that the complete
condensation takes place for any $\rho>0$ on random networks by
measuring average mass of a node with degree $k$. Hence the
complete condensation always takes place for any nonzero density
in ZRP with constant chipping rate on random and scale-free
networks.

Finally we investigate the behavior of the  average mass of a node
with degree $k$, $m_k$ in the fluid and the condensed phase. In
ZRP with constant chipping rate, $m_k$ linearly increases with
degree $k$, and jumps to the total mass of the system at hub
degree $k_{hub}$ \cite{NSL}. Hence in SCA model, $m_k$ is expected
to show such a jump in the condensed phase for $\omega \neq
\infty$. However $m_k$ linearly increases up to $k_{hub}$ without
jumps in both the fluid and the condensed phase. We numerically
confirm the linearity of $m_k$, and also analytically show $m_k
\sim k$ with the assumption that the diffusion is the only
relevant factor in the steady state. Therefore the steady state
distribution of $m_k$ is determined by diffusion of masses rather
than chipping of unit mass unlike in ZRP.

This work is supported by Grant No. R01-2004-000-10148-0 from the
Basic Research Program of KOSEF. We also thank Prof. Soon-Hyung
Yook for critical reading of our manuscript.

\newpage
\begin{figure}
\includegraphics[width=8.5cm]{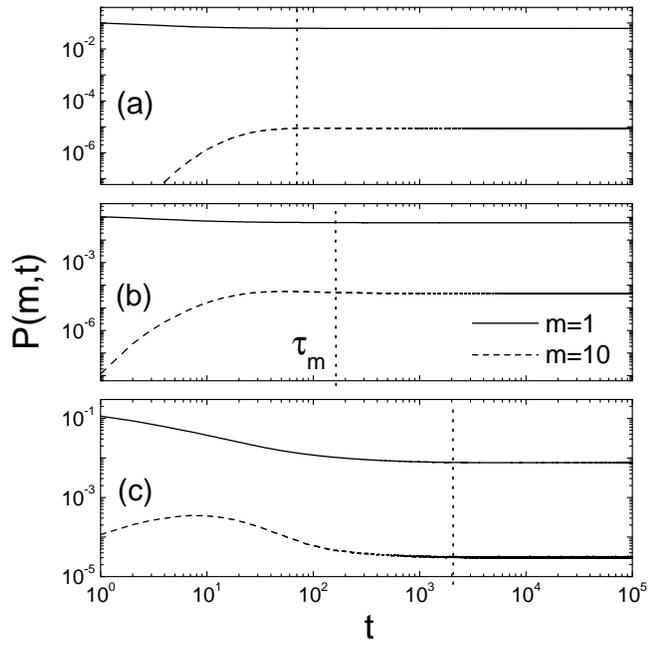}
\caption{$P(m,t)$ of $m=1$ and $10$ for $\rho=0.1$ and $\omega=1$
on networks of $N=10^4$. Each panel corresponds to RN (a), SFN
with $\gamma=4.3$ (b) and $\gamma=2.4$ (c). The vertical dotted
line denotes the characteristic time $\tau_m$.}
\end{figure}

\begin{figure}
\includegraphics[width=8.5cm]{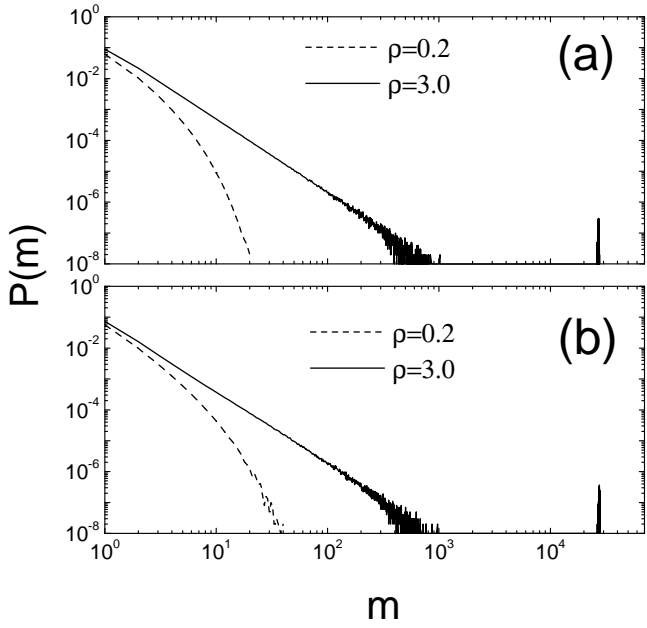}
\caption{$P(m)$ on RN (a) and SFN with $\gamma=4.3$ (b) at
$\omega=1.0$. In each panel, the solid and the dashed line
correspond to $P(m)$ of $\rho = 3.0$ (condensed phase) and $0.2$
(fluid phase) respectively. }
\end{figure}

\begin{figure}
\includegraphics[width=8.5cm]{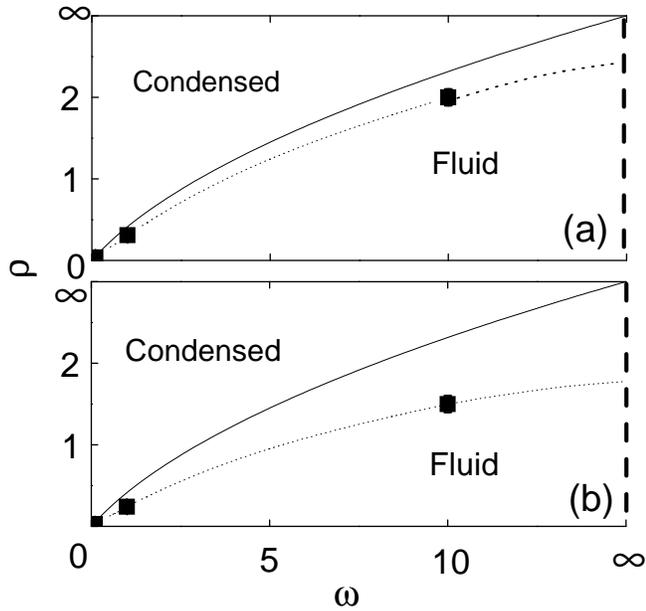}
\caption{ The $\rho$-$\omega$ phase diagram of RN (a) and SFN with
$\gamma=4.3$ (b). In each panel, the solid line is the critical
line of $\rho_c (\omega) = \sqrt{\omega+1} -1$ on regular lattices
of Ref. \cite{MKB}. The dotted line between data points is a guide
to the eye. The dashed line denotes $\omega = \infty$ line. }
\end{figure}

\begin{figure}
\includegraphics[width=8cm]{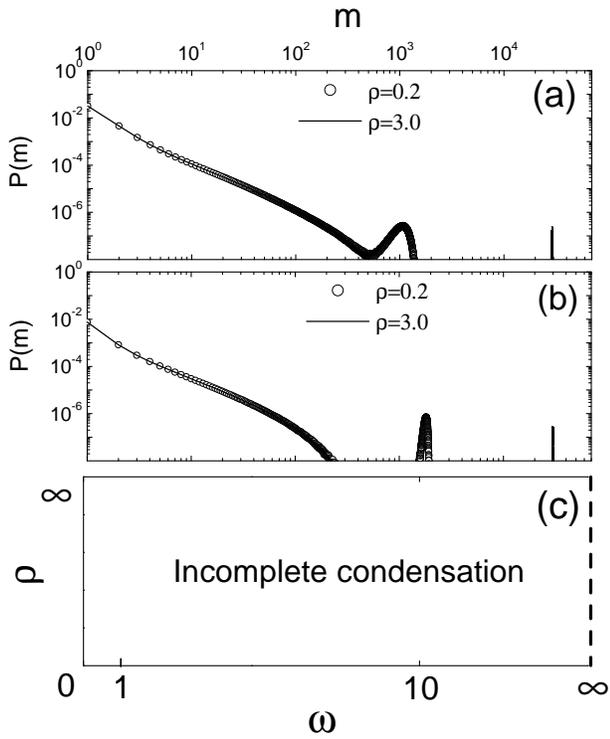}
\caption{ $P(m)$ on SFN with $\gamma=3$ (a) and $\gamma=2.4$ (b)
at $\omega=1.0$. In each panel, the solid and the dashed line
correspond to $P(m)$ of $\rho = 3.0$ and $0.2$ respectively. (c)
the typical phase diagram of SFNs with $\gamma \leq 3$. The dashed
line denotes the line of $\omega = \infty$. }
\end{figure}

\begin{figure}[t]
\includegraphics[width=8.5cm]{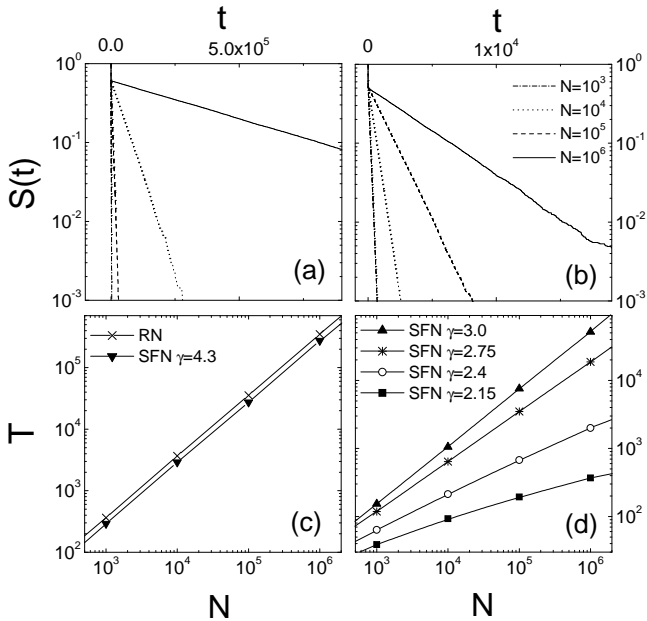}
\caption{Semi-log plot of the survival probability $S(t)$ on SFN
with $\gamma=4.3$ (a) and $2.4$ (b) of various size $N$ up to
$10^6$. (c) The average life time $T$ on RN and SFN with
$\gamma=4.3$. (d) $T$ on SFNs with $\gamma \leq 3$. The solid line
between data points is a guide to the eye.}
\end{figure}

\begin{figure}
\includegraphics[width=8.5cm]{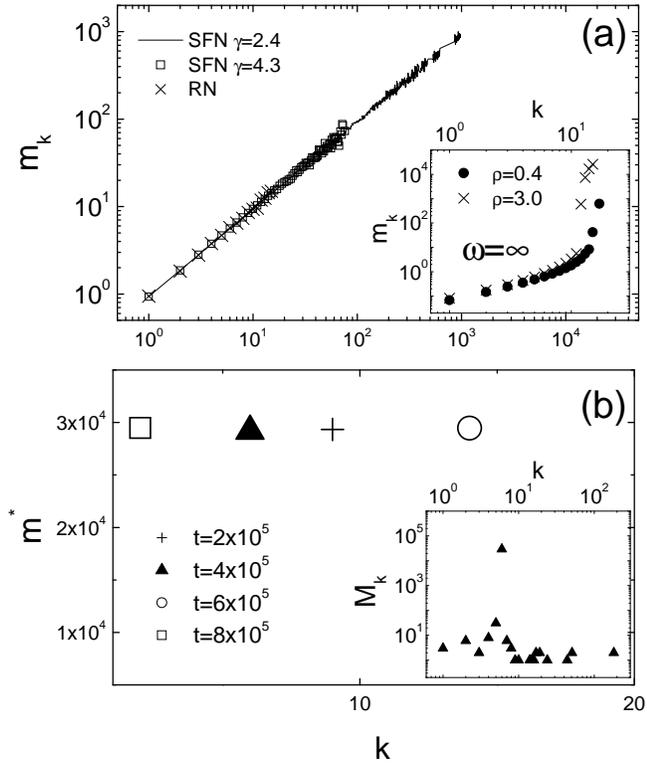}
\caption{(a) The main plot shows $m_k$ of Eq. (10) for $\rho=3.0$
and $\omega=1$ on RN, SFN with $\gamma=4.3$ and $2.4$ in condensed
phase. The inset shows $m_k$ at $\omega = \infty$ on RN. (b) $M_k$
of a single sample on SFN with $\gamma=2.4$ at different time for
$\rho =3$ and $\omega=1$ of the condensed phase. The main plot
only shows the maximum mass $m^*$ of $M_k$ of a single sample at
different time steps, and the inset shows the snapshot of the
$M_k$ at $t=4\times10^5$. }
\end{figure}
\end{document}